\documentclass[manuscript]{emulateapj}

\slugcomment{Submitted to ApJ}

\shorttitle{ M31N 2007-12b: An Extragalactic Recurrent Nova?}
\shortauthors{Bode et al.}

\begin{document}

\title{Optical and X-ray Observations of M31N 2007-12\lowercase{b}: An Extragalactic Recurrent Nova with a Detected Progenitor?}

\author{M. F. Bode and M. J. Darnley}
\affil{Astrophysics Research Institute, Liverpool John Moores University, Birkenhead, CH41 1LD, UK}

\author{A.W. Shafter}
\affil{Department of Astronomy, San Diego State University, San Diego, CA 92182}

\author{K. L. Page}
\affil{Department of Physics and Astronomy, University of Leicester, Leicester, LE1 7RH, UK}

\author{O. Smirnova}
\affil{ Institute of Astronomy, University of Latvia, Raina Boulevard 19, LV-1586 Riga, Latvia}

\author{G.C. Anupama}
\affil{Indian Institute of Astrophysics, Koromangala, Bangalore 560 034, India }

\and

\author{T. Hilton}
\affil{Astrophysics Research Institute, Liverpool John Moores University, Birkenhead, CH41 1LD, UK}

\begin{abstract}

We report combined optical and X-ray observations of nova M31N
2007-12b.  Optical spectroscopy obtained 5 days after the 2007
December outburst shows evidence of very high ejection velocities
(FWHM H$\alpha \simeq 4500$ km s$^{-1}$). In addition, $Swift$ X-ray
data show that M31N 2007-12b is associated with a Super-Soft Source
(SSS) which appeared between 21 and 35 days post-outburst and turned
off between then and day 169. Our analysis implies that $M_{\rm WD}
\ga 1.3~$M$_{\odot}$ in this system. The optical light curve, spectrum
and X-ray behaviour are consistent with those of a recurrent nova. {\it Hubble Space Telescope} observations of the pre-outburst
location of M31N 2007-12b reveal the presence of a coincident stellar
source with magnitude and color very similar to the Galactic recurrent nova
RS Ophiuchi at quiescence, where the red giant secondary dominates the
emission. We believe that this is the first occasion on which a nova
progenitor system has been identified in M31. However, the greatest
similarities of outburst optical spectrum and SSS behaviour are with
the supposed Galactic recurrent nova V2491 Cygni. A previously implied association of M31N
2007-12b with nova M31N 1969-08a is shown to be erroneous and this has
important lessons for future searches for recurrent novae in extragalactic
systems. Overall, we show that suitable complementary X-ray and
optical observations can be used not only to identify recurrent nova
candidates in M31, but also to determine subtypes and important
physical parameters of these systems. Prospects are therefore good for
extending studies of recurrent novae into the Local Group with the potential to
explore in more detail such important topics as their proposed link to
Type Ia Supernovae.

\end{abstract}


\section{Introduction}

Classical Novae (CNe) are cataclysmic variable stars whose outbursts
are due to a Thermonuclear Runaway (TNR) on the surface of a white
dwarf in an interacting binary system \citep[see
  e.g.][]{sta08}. Recurrent Novae (RNe) are related to CNe, but have
been seen to undergo more than one recorded outburst and may contain
evolved secondary (mass-donating) stars \citep[see][for a
  review]{anu08}. Recurrent novae have been proposed as one of the
primary candidates for the progenitors of Type Ia Supernovae
\citep[SNe - see e.g.][for a recent review]{kot08}.

At present we know of a total of only 10 RNe in the Galaxy with confidence (based on two or more nova outbursts being observed). These RNe 
appear to fall into three main groups, {\em viz.}: \\{\em RS Oph/T
  CrB} with red giant secondaries, consequent long orbital periods
($\sim$ several hundred days), rapid declines from outburst ($\sim
0.3$ mag day$^{-1}$), high initial ejection velocities ($\ga 4000$ km
s$^{-1}$) and strong evidence of the interaction of the ejecta with
the pre-existing circumstellar wind of the red giant \citep[from
  observations of optical coronal lines, non-thermal radio emission
  and hard X-ray development in RS Oph; see papers in][]{eb08};\\  The
more heterogeneous {\em U Sco} group with members' central systems
containing an evolved main sequence or sub-giant secondary with an
orbital period much more similar to that in CNe (of order hours to a
day), rapid optical declines (U Sco itself being one of the fastest
declining novae of any type), extremely high ejection velocities
($v_{ej} \sim 10,000$ km s$^{-1}$, from FWZI of emission lines for U
Sco) but no evidence of the extent of shock interactions seen in RS Oph
post-outburst \citep[their post-outburst optical spectra resemble the
  `He/N' class of CNe --][]{wil92};\\ {\em T Pyx, CI Aql, IM Nor} are
again short orbital period systems and although their optical spectral
evolution post-outburst is similar, with their early time spectra
resembling the `Fe II' CNe, they show a very heterogenous set of
moderately fast to slow declines in their optical light curves. This
sub-group of RNe also seems to show ejected masses similar to
those at the lower end of the ejected mass range for CNe with $M_{\rm
  ej} \sim 10^{-5}$ M$_{\odot}$ (which appears to be one to two orders of magnitude
greater than $M_{\rm ej} $ in the other two sub-groups of RNe).

The short recurrence periods of RNe require high mass WD accretors and
relatively high accretion rates \citep[e.g. ][]{sta88}. Indeed, both
RS Oph and U Sco appear to have WDs near to the Chandrasekhar mass
limit. The WD mass in both these objects has been proposed as growing
and therefore they are potential SN Ia progenitors \citep[see
  e.g.][respectively]{sok06,kah99}

The study of RNe is thus important for several broader fields of
investigation including mass loss from red giants, the evolution of
supernova remnants and the progenitors of Type Ia SNe. Progress in
determining the latter association in particular, as well as exploring
the evolutionary history of these close binary systems, is hampered by
the relative rarity of Galactic RNe. However, since the time of Edwin
Hubble, CNe have been observed in extragalactic systems, in particular
M31 \citep[see][for a review]{sha08}. In total over 800 CN candidates
have been catalogued in M31 \citep{pie07a} and among these are thought
to lie several RNe \citep[see e.g.][]{del96, sha09b}. Indeed
\cite{pie07a} identified 4 candidates in their search for the X-ray
counterparts of optical novae in M31 \citep[see also][]{hen09}. In
this paper we present evidence for an object in M31 previously
classified as being a CN as in fact being a recurrent nova. We use a
combination of optical and X-ray observations to explore its more
detailed nature, emphasise the need for careful exploration of
archival material to confirm or rule out previous outbursts, and go on
to point the way to more extensive observational programs in the
future.

\section{Observations of the 2007 Outburst}

\begin{figure}
\includegraphics[angle=270,width=\textwidth]{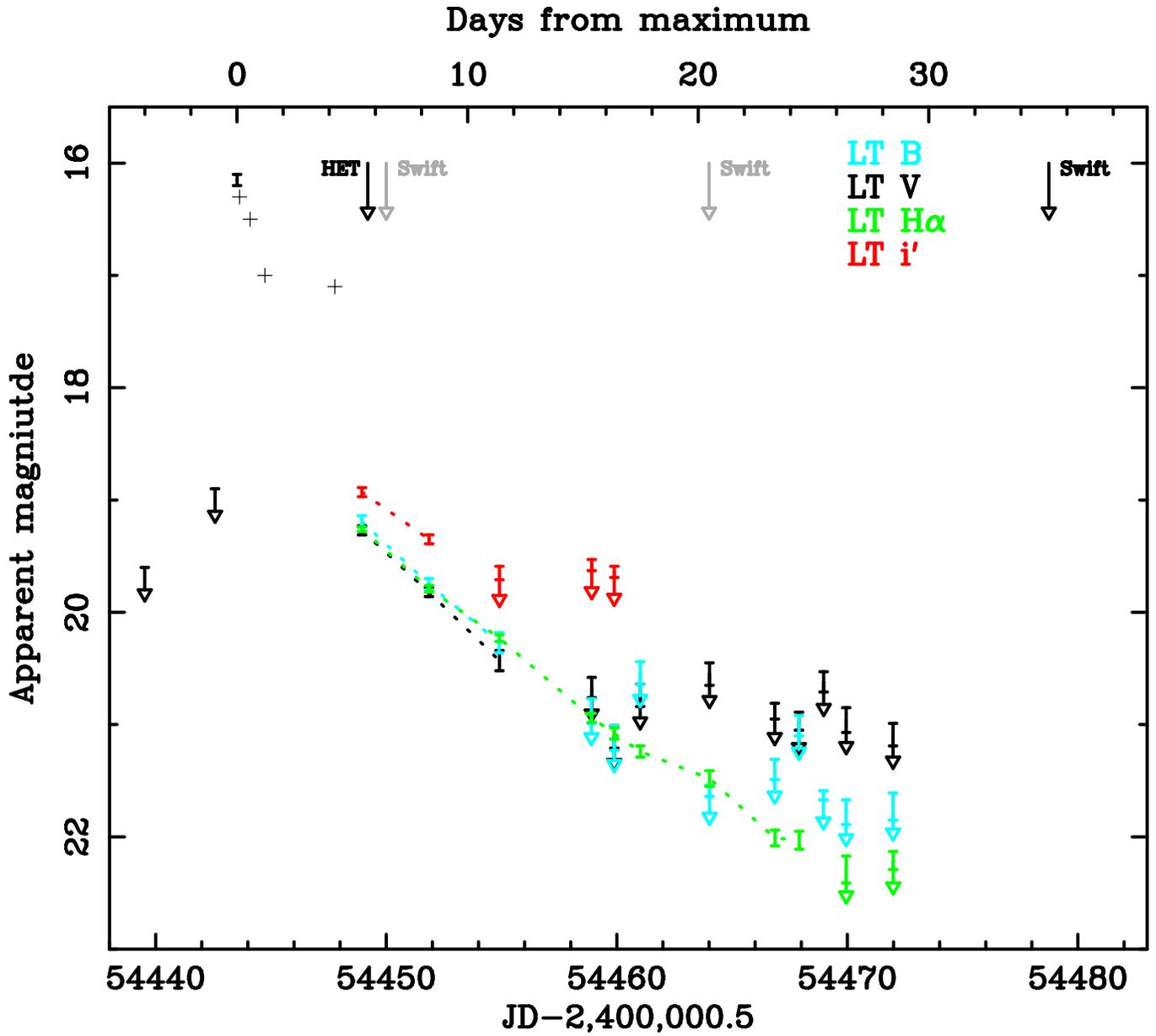}
\caption{The optical light curve of M31N 2007-12b with data from \cite{lee07} and the ``M31 (Apparent) Nova Page'' provided by the International Astronomical Union, Central Bureau for Astronomical Telegrams (CBAT - http://www.cfa.harvard.edu/iau/CBAT\_M31.html) comprising both $R$ band and unfiltered CCD observations, plus data from the Liverpool Telescope (data points from $t \sim 5$ days after peak and in photometric bands as indicated). Times of {\em Swift} observations (upper limits grey; detection at $t = 35$ days bold) and HET spectroscopy are also shown.}\label{fig1}
\end{figure}

Nova M31N 2007-12b was discovered on 2007 December 9.53 UT (which we
take as $t = 0$) by K. Nishiyama and
F. Kabashima\footnote{http://www.cfa.harvard.edu/iau/CBAT\_M31.html}
at mag $= 16.1-16.2$ (unfiltered)  and located at RA = 00h 43m
19s.94$\pm$0s.01, DECL $= +41\degr 13' 46''.6 \pm 0''.1$ (J2000). They
reported that no object had been visible at this position on 2007
December 8.574 UT. Fig.\ref{fig1} gives details of these and other
optical observations around peak. Broadband $i', V, B$ plus narrowband
H$\alpha$ photometry was subsequently obtained with the RATCam CCD
camera  on the 2-m Liverpool Telescope \citep[LT; ][]{ste04}. LT
photometry is part of a larger program of photometry and spectroscopy
of novae in M31 \citep{sha09} and began on 2007 December 14.94 UT ($t
= 5.4$ days post-outburst) then continued for 23 days.  The LT data
were reduced using standard routines within the IRAF\footnote{IRAF is
  distributed by the National Optical Astronomy Observatory, which is
  operated by the Association for Research in Astronomy, Inc. under
  cooperative agreement with the National Science Foundation.} and
STARLINK packages, and calibrated against standard stars from
\cite{lan92} and by using the secondary standards in M31
\citep{mag92,hai94}. The resulting lightcurves are shown in
Fig.\ref{fig1}.

The astrometric position of M31N 2007-12b was measured from an LT Sloan
$i'$-band image taken on 2007 December 14.95 UT.  This image was chosen as
a compromise between good seeing and nova brightness.  An
astrometric solution was obtained using 21 stars from the 2MASS
All-Sky Catalogue \citep{2003tmc..book.....C} which are coincident with
resolved sources in the LT observation.  We obtain a position for M31N
2007-12b of RA = 00h 43m 19s.97$\pm$0s.01 DECL $= +41\degr 13' 46''.3
\pm 0''.1$ (J2000; consistent with Nishiyama and Kabashima's measurement).  It should be
noted that the astrometric uncertainty is dominated by uncertainties
in the plate solution.

Optical spectroscopy was obtained by us on 2007 December 15.2 UT ($t =
5.7$ days) with the 9.2-m Hobby Eberly Telescope (HET) using the Low
Resolution Spectrograph \citep [LRS;][]{hil98}. We used the $g$1
grating with a $1.0''$ slit and the GG385 blocking filter, which
covers $4150-11000$ \AA ~with a resolution of $R \sim 300$, although
we limit any analysis to the $4150-8900$ \AA ~range where the
effects of order overlap are minimal. Data reduction was performed
using standard IRAF packages and the resulting spectrum is shown in
Fig.\ref{fig2}.

\cite{kon07} reported the detection of a Super-Soft X-ray Source (SSS)
coincident with the position of the nova using the X-ray Telescope
(XRT) on board the $Swift$ satellite \citep{bur05}. The detection was
made serendipitously as part of a survey of SSSs in the M31 globular
cluster Bol 194 on 2008 January 13.74 UT ($t = 35.2$ days) with an
exposure time of 4 ks. They reported previous observations of the
field on 2007 December 16 and December 30 that had not detected any
source at that position. We have re-analysed the XRT data for these
epochs and also consulted the {\em Swift} data archive to review other X-ray
observations of this field from 2007 November to 2008 May (see Table
\ref{xrt} and also Fig.\ref{fig1}).

\begin{table}
\begin{center}
\caption{~{\em Swift} XRT data. Upper limits are at the 90\% confidence level; error on the detection is $1 \sigma$.}
\begin{tabular}{lccc}
\hline \hline
Date (day) & Obs ID & Exposure  & count rate\\
           &        & time (ks) & (s$^{-1}$) \\
\hline

2007-11-24 (-15)& 00031027001 & 7.27 & $<$0.0017 \\
2007-12-02 (-7)& 00031027002 & 1.00 & $<$0.0073 \\
2007-12-03 (-6)& 00031027003 & 3.63 & $<$0.0037\\
2007-12-16 (+7)& 00031027004 & 3.89 & $<$0.0039 \\
2007-12-30 (+21)& 00031027005 & 4.02 & $<$0.0034 \\
2008-01-13 (+35)& 00031027006 & 3.99 & 0.015 $\pm$ 0.002\\
2008-05-26 (+169)& 00037719001 & 4.86 & $<$0.0023\\

\hline
\end{tabular}
\label{xrt}
\end{center}
\end{table}

\section{Results and Discussion}

Nova M31N 2007-12b lies within $1.7''$ of the quoted position of M31N
1969-08a (RA  = 00h 43m 19s.9 $\pm 0s.3$, DECL $ = +41\degr 13' 45''
\pm3''$ (J2000); i.e. coincident within the quoted measurement errors)
which was discovered on 1969 August 16.0 UT \citep[see][]{sha91} and
lies $7.1'$ from the nucleus of M31. Peak visual magnitude was
observed one day after the start of the 1969 outburst at $V =
16.4$. Subsequently, the nova declined at a rate of $\ga 0.3$ mag
day$^{-1}$ making this a very fast nova \citep{war08}. Supposed
positional coincidence and similarities in their light curves led to the initial conclusion that the
outbursts were from the same object. However, consultation of the
original plate material for M31N 1969-08a showed that its position is
in fact RA  = 00h 43m 19s.6 $\pm0s.1$, DECL $ = +41\degr 13' 44''\pm1'' $ (J2000, i.e. separated by $4.8''\pm1.5''$ from M31 2007-12b) and blinking of the 1969
and 2007 images confirmed they are indeed separate objects (see Fig.\ref{finder}).

\begin{figure}
\includegraphics[angle=270,width=\textwidth]{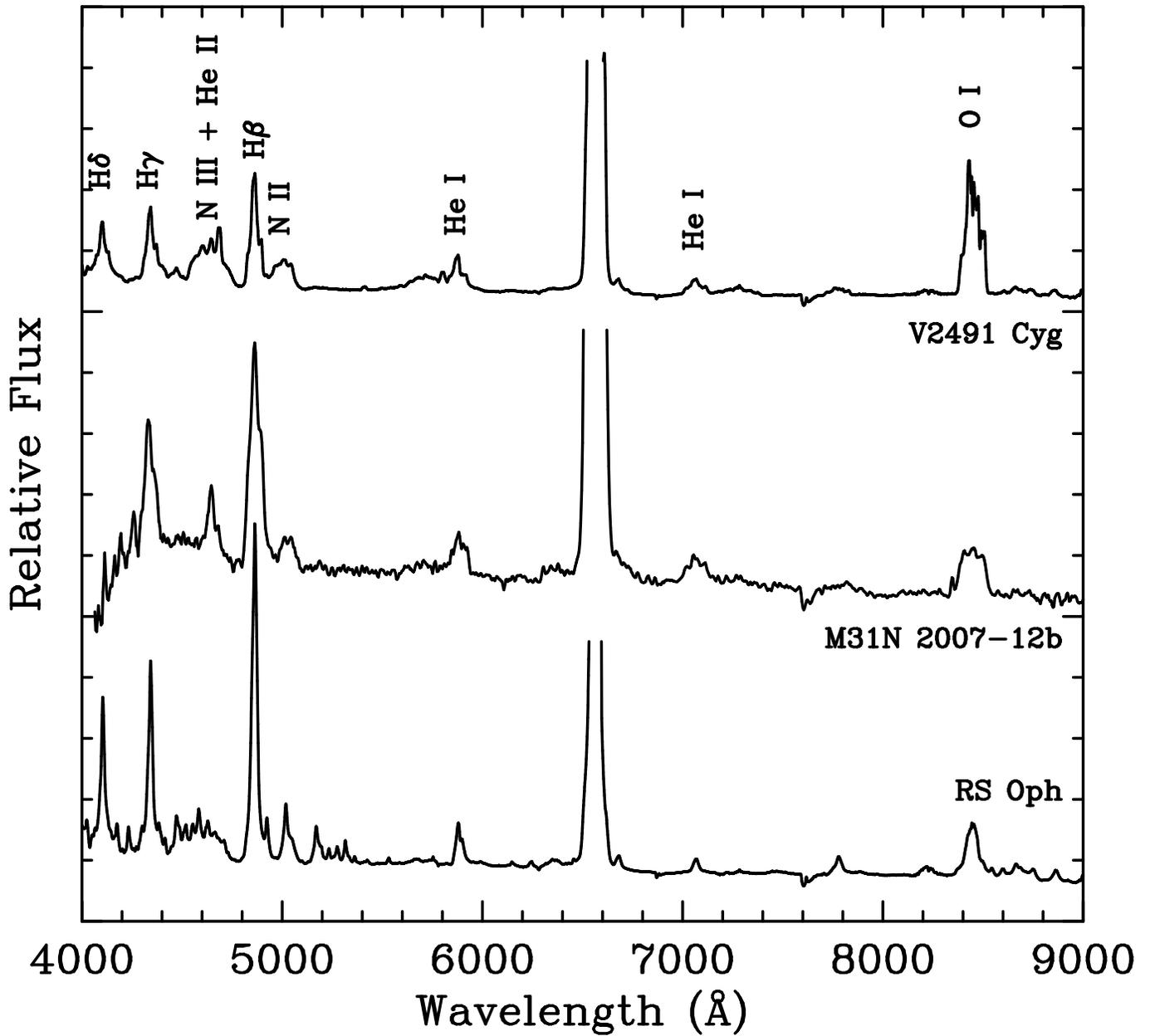}
\caption{Low resolution spectrum of M31N 2007-12b taken with the LRS on the HET at $t = 5.2$ days after discovery.  The spectrum would classify this object as a He/N Classical Nova following maximum light. Also shown for comparison are spectra of the Recurrent Nova RS Oph at around 6 days from outburst \citep{anu08} and the supposed RN V2491 Cyg (from two co-added spectra at around 17 and 18 days after outburst, Anupama et al. in preparation) - see text for further details.  The H$\alpha$ line in each spectrum has been truncated.\label{fig2}}
\end{figure}

\begin{figure}
  \includegraphics[width=\textwidth]{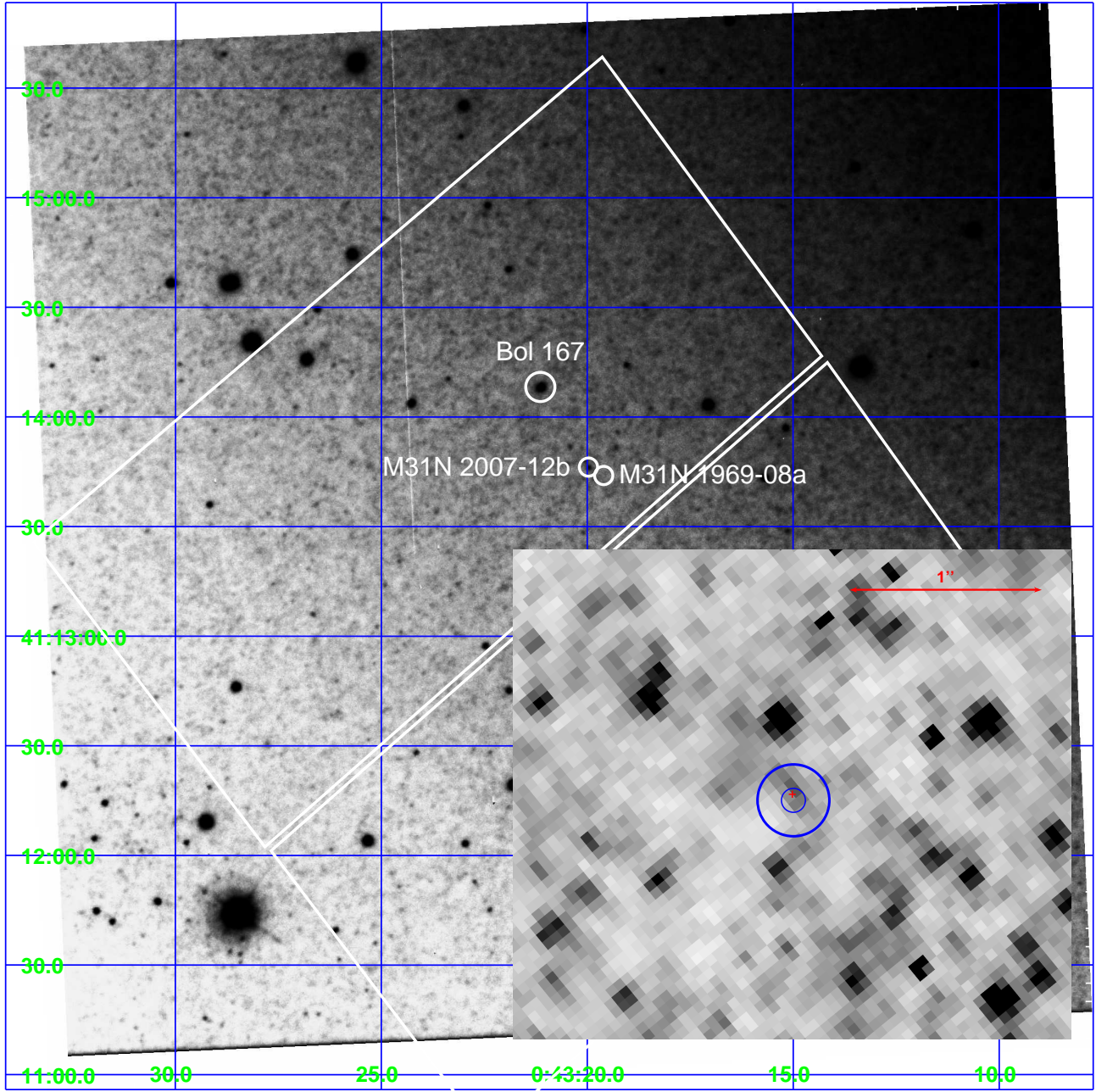}
\caption{Main image: Liverpool Telescope Sloan-$i'$ image of M31N 2007-12b taken on 2007 December 14.95 UT, field $4.6'\times4.6'$.  The position of M31N 2007-12b and the new position of M31N 1969-08a are shown by the small white circles (left and right respectively, circles have a diameter of $5''$).  The white boxes indicate the approximate positions of the {\em HST} ACS-WFC field.  Also shown is the position of the nearby globular cluster Bol 167.  Inset: HST ACS-WFC image of the $\sim3''\times3''$ region surrounding M31N 2007-12b. The inner blue circle indicates the $1\sigma$ ($1.25''$) radius search region for the progenitor, the outer circle the $3\sigma$ region, the red cross indicates the position of the progenitor candidate.\label{finder}}
\end{figure}

Our optical spectroscopy on day 5.7, shown in Fig. \ref{fig2}, reveals
strong and very broad (FWHM H$\alpha \simeq 4500$ km s$^{-1}$) Balmer,
He I and N III 464.0 nm emission lines consistent with the spectra of
He/N CNe \citep{wil92}. High emission line velocities and fast optical
declines are associated with ejection from a high mass WD and are also
typical of both the RS Oph and U Sco sub-classes of RNe
\citep{anu08}. Of these two, the spectrum more closely resembles that
of RS Oph around 3 days after the 2006 outburst (see Fig. \ref{fig2}),
than that observed in U Sco or the U Sco sub-class RN V394 CrA at
similar phases after their outbursts in 1987
\citep{sek88,sek89,wil91}. However, the most striking spectral
similarity is to the early optical spectrum of nova V2491 Cyg (again,
see Fig. \ref{fig2}) for which, although only one outburst has been
observed, it has been suggested that it is a RN \citep{pag09} by
virtue of its very fast optical decay and high ejection velocities
together with its low outburst amplitude \citep[$\Delta V = 8.5$
  mag,][]{jur08} and detection as an X-ray source pre-outburst
\citep{iba09}.

\subsection{Constraints from the X-ray data}

Turning now to the X-ray spectra, we re-analysed the {\em Swift} detection
on day 35.2 referenced by \cite{kon07} using the {\em Swift} software
version 2.9. Source spectra were extracted from the cleaned Photon
Counting mode event lists, using a 10 pixel extraction radius (1 pixel
$= 2.36''$). A total of 49 background-subtracted counts were found
with only one count at $\sim 0.9$ keV and the rest at lower
energies. As an initial guide this super-soft spectrum was then fitted
with an absorbed black body spectrum using XSPEC. We estimated the
absorbing column as follows. \cite{sta92} derive a Galactic
contribution to the column in this direction equivalent to $E_{B-V}$ =
0.1. At the position of M31N 2007-12b in M31, following the
methodology discussed in \cite{dar06} and Section 3.2 below, and
assuming that the nova is situated half way down the absorbing column
internal to M31, we get $E_{B-V}$ = 0.25. Thus the total extinction to
the nova is estimated to be equivalent to $E_{B-V}$ = 0.35, which in
turn is equivalent to $N_{\rm H} = 2.1\times10^{21}$ cm$^{-2}$. The
best fit to the data using this total column then gives $kT =
63^{+10}_{-8}$ eV (i.e. $T = 7.6\pm{1.2} \times 10^5$K) and for $d =
780$ kpc to M31 \citep{hol98, sta98} yields an absorption-corrected
luminosity $L = (4.5^{+1.9}_{-1.4}) \times10^{38}$ ergs s$^{-1}$
(i.e. around twice the Eddington luminosity for a 1.4 M$_\odot$
WD). We obtained non-detections at the source position in 2007
November/December and 2008 May as detailed in Table \ref{xrt}.

From the above observations with {\em Swift}, the SSS was not detected $\sim
15$, 7 and 6 days before outburst and at 7, 21 and 169 days
afterwards.  We can estimate therefore that the SSS appeared between
$t \sim 21$ and 35 days post-outburst and had turned off again at $t <
169$ days. A caveat here is that the onset of the SSS phase has shown extreme variability in a few objects so far \citep[e.g. RS Oph; see][]{pag08} and there is the possibility that the first emergence was earlier than 21 days.
We can however compare the observed behavior of the M31N 2007-12b SSS with the properties of this phase in possibly
related Galactic novae. For example, we note that the appearance of
the SSS in U Sco was around 19-20 days after the peak of the optical outburst in February 1999 \citep{kah99}. 
We used the model parameters for U Sco found by \cite{kah99} to generate a spectrum with the correct unabsorbed flux. In order to determine the predicted count rate in the {\em Swift} XRT if the source were placed in M31 at $d = 780$ kpc, the absorbing column was changed to $N_{\rm H} =  2.1\times10^{21}$ cm$^{-2}$ but the normalization and derived $kT$ were kept fixed. Finally, a new spectrum was generated to derive  
the predicted count-rate. Table \ref{novae3} gives details of the parameters of this and other sources described below, together with their derived count rates.

It can be seen from Table \ref{novae3} that with the
spectral parameters given in \cite{kah99}, the SSS emission
seen in U Sco ($d = 14$ kpc) would not have been
detectable by $Swift$ at the distance of M31 in the exposure times used
for M31N 2007-12b, although of course the U Sco SSS may have
subsequently increased in brightness. In RS Oph ($d =1.6$ kpc) the SSS
emerged and then dominated the X-ray emission from $t \sim 29$ days and 
turned off by $\sim 90$ days \citep{pag08}, i.e. consistent with the
timescales in M31N 2007-12b. However, as can be seen from Table \ref{novae3}, even at the peak of its SSS
emission, RS Oph would also have been undetected in M31 with
the $Swift$ XRT. In V2491 Cyg ($d =
10.5$ kpc), SSS emission became apparent after around 25 days
\citep{pag09} and was sharply peaked at around 40 days. At the
distance of M31, the {\em Swift} XRT observations reported here would have
detected the V2491 Cyg SSS for a few days around this peak (again, see Table \ref{novae3}).

\begin{table*}
\begin{center}
\caption{~Parameters used to derive predicted Swift count rates for the SSS phase in other novae if at the distance and absorbing column of M31N 2007-12b (see text for details).}
\begin{tabular}{lcccccl}
\hline \hline
& $d$ & $N_{\rm H}$  & kT$_{BB}$ & Unabsorbed Flux & Predicted Count Rate \\
& (kpc) & (cm$^{-2}$) & (eV) & (ergs s$^{-1}$ cm$^{-2}$) & (s$^{-1}$) \\
\hline
U Sco$^{\rm a}$ & 14 & $2.2\times10^{21}$ & 107 & $5.4\times10^{-10}$ & $1\times10^{-3}$ \\
RS Oph$^{\rm b}$ & 1.6 & $3.4\times10^{21}$ & 70 & $6.3\times10^{-8}$ & $1.1\times10^{-3}$ \\
V2491 Cyg$^{\rm c}$ & 10.5 &  $3.4\times10^{21}$ & 52 & $2.4\times10^{-8}$ & $9.4\times10^{-3}$ \\
\hline
\end{tabular}
\label{novae3}
\end{center}
$^{\rm a}$ From \cite{kah99} and assuming the unabsorbed flux they quote is for the 0.1-10 keV energy range of the LECS/MECS of {\em BeppoSAX}.\\
$^{\rm b}$ Fit to the {\em Swift} XRT data from day 50.5 after outburst during the SSS `plateau' phase. The unabsorbed flux is for the 0.3-10 keV energy range of the XRT. The $N_{\rm H}$ value used in the fit includes both interstellar and circumstellar components \citep[see][]{pag08}.\\
$^{\rm c}$ Fit to the {\em Swift} XRT data from day 41.7 after outburst around the observed SSS peak count rate \citep{pag09}. The unabsorbed flux is for the 0.3-10 keV energy range of the XRT.
\end{table*}


It is well established that the SSS arises from continued nuclear
burning on the WD surface following the TNR which is gradually
unveiled as the ejecta move outwards \citep{kra08}. The deduced
temperature and luminosity of the SSS in the case of M31N 2007-12b are
consistent with this model. Simplistically, the timescale for
uncovering and observed onset of the SSS phase is given by $t_{on}
\propto M_{\rm H}^{1/2} v_{ej}^{-1}$ \citep{kra96} where $M_{\rm H}$
is the mass of H in the ejected envelope and $v_{ej}$ is the ejection
velocity. Thus for the low ejected masses and high ejecta velocities
found in  RS Oph-type and U Sco-type RNe, $t_{on}$ would be expected
to be relatively short compared to that for the T Pyx sub-class of RNe or its value for most CNe.  

The turn-off time since outburst for nuclear burning, $t_{\rm rem}$,
is a steep function of WD mass. \cite{mac96} finds for example
$t_{\rm rem} \propto M_{\rm WD}^{-6.3}$. Generally in CNe this
timescale is much longer than that observed in M31N 2007-12b
\citep[although][make the point that this might be ascribed in part to
  a selection effect]{pie07a}. For example, in one of the best studied
cases, the moderately fast CN  V1974 Cyg, $511 < t_{\rm rem} < 612$
days \citep{bal98} with $M_{\rm WD} \sim 1$ M$_\odot$
\citep{hac06}. From \cite{sta91}, $t_{\rm rem} < 169$ days implies
$M_{\rm WD} \ga 1.3~$M$_{\odot}$. Similarly, the timescale after
outburst for the onset of the SSS phase, $t_{\rm on}$, is also a
function of $M_{\rm WD}$ in the sense that $t_{\rm on}$ is likely to
be shorter for systems containing a high mass WD. As noted above, both
U Sco and RS Oph have very short observed $t_{\rm on}$ of $\lesssim
20$ days \citep{kah99} and $\lesssim 30$ days \citep{bod06}
respectively. In both cases, the WD mass is determined to be
approaching the Chandrasekhar mass limit of M$_{\rm Ch} \sim
1.4$ M$_{\odot}$ \citep{kah99,hac07}. Similarly, the WD mass in V2491
Cyg is estimated to be $> 1.3$ M$_{\odot}$ \citep{hac09,pag09}. In
addition, we note that envelope composition may also be important in
determining the duration of the SSS phase. However, with  $t_{on} <
35$ days and $t_{rem} < 169$ days, $M_{\rm WD} > 1.3$ M$_\odot$ for
the range of envelope compositions presented in \cite{hac06} and in
particular,  $M_{\rm WD} > 1.35$ M$_\odot$ for the cases of (fast)
Neon novae they present.


\subsection{A search for the progenitor system}

\begin{figure}
\includegraphics[angle=270,width=\columnwidth]{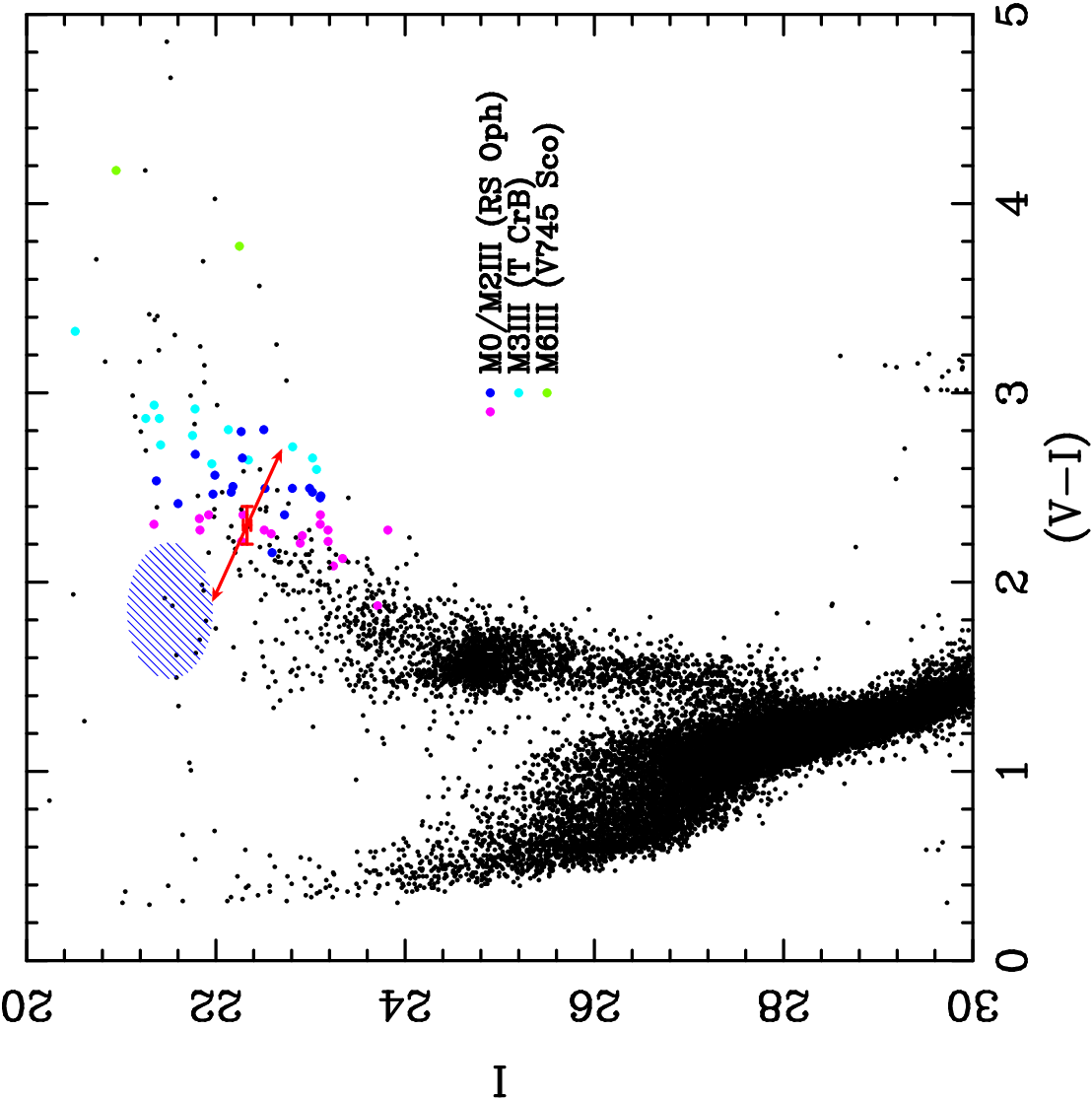}
\caption{Color-magnitude diagram showing the {\it Hipparcos}
  data-set \citep{1997ESASP1200.....P}, with parallax and photometric
  errors $<10\%$.  The {\it Hipparcos} stars have been moved to the
  position of M31 assuming $(m-M)_{0}=24.43$
    \citep{1990ApJ...365..186F} an estimated extinction towards M31
    of $E_{B-V}=0.1$\ mag \citep{sta92} and an
  internal extinction of $E_{B-V}=0.25$.  The red
  point shows the candidate for the progenitor of M31N 2007-12b. The colored points are known M0/M2III (RS Oph), M3III (T CrB)
  and M6III (V745 Sco) secondary stars of RNe \protect{\citep{anu08}}.  The dark blue hatched ellipse
  shows the location of a quiescent RS Oph system (with the major and minor axes of the ellipse relating to $1\sigma$ scatter from the mean). The red arrow shows the extent of the likely effects of internal extinction ($\Delta E_{B-V}=\pm0.25$).\label{fig4}}
\end{figure}

If M31N 2007-12b arose from a RN system of the RS Oph sub-type, it
would contain a red giant secondary.  We thus explored its detection
at quiescence in archival {\it Hubble Space Telescope (HST)}
imagery. The  {\it HST} is capable of resolving giant branch stars
within M31 (see Fig.\ref{fig4}).  The positions of both M31N
2007-12b and M31N 1969-08a lie within a pair of archival {\it HST}
Advanced Camera for Surveys (ACS) Wide Field Channel (WFC) images
(prop. ID 10273) taken in August 2004 using the F814W ($\sim I$) and
the F555W ($\sim V$) filters.  PSF fitting photometry was performed on
all detected objects in both {\it HST} pass-bands using DOLPHOT, a
photometry package based on HSTphot \citep{2000PASP..112.1383D}.  We
used the relations given in \cite{2005PASP..117.1049S} to transform
from these filters to Johnson-Cousins $V$ and $I$.

To isolate the position of M31N 2007-12b within the {\it HST} data we
computed the spatial transformation between the {\it LT} and Gaussian
convolved {\it HST} data using 23 stars resolved and unsaturated in
both images.  This approach is independent of the astrometric calibration of both fields and hence yields the most accurate results.  The uncertainty in the derived transformation is small
when compared to the 0.22 pixel ($0.06^{\prime\prime}$) average positional
error of the nova in the {\it LT} data.  This positional uncertainty
in the {\it LT} data equates to a 1.25 pixel positional uncertainty
($1\sigma$) within the {\it HST} data.  

There is a resolved object just inside $1\sigma$ from the {\it LT}
position (separated 1.12 {\it HST} pixels or $0.89\sigma$) seen in the
{\it HST} F555W image (see inset of Fig.\ref{finder}).  We find that this object has $V=24.61\pm0.09$
and $I=22.33\pm0.04$, hence a color of $V-I=2.3\pm0.1$.  It should be
noted that there is a cosmic ray track very close to this object's position
in the F814W image, hence the $I$-band photometry may have been
adversely affected by the subtraction of the cosmic ray.  There are no
other resolved stars within 1.90 {\it HST} pixels or $1.52\sigma$.

Shown in Fig.\ref{fig4} is the position on a color-magnitude diagram of the object spatially coincident with M31N
2007-12b.  This object (assuming no additional internal M31 extinction)
lies in the M0/M2III (RS Oph secondary, purple and dark blue dots respectively) and M3III (T CrB secondary, light blue dots) region of the
Giant Branch.  The probability of finding such a star ($20<I<23$,
$1.5<V-I<2.5$) at least as close to the predicted position by chance
is only 3.4\%. We note as an aside that we have explored the region around M31N 1969-08a and found no significant spatial coincidence with any pre-existing stellar source. 

We estimate the mean $I$-band extinction across an Sb galaxy, such as M31,
to be $A(I)=0.8$ magnitudes, equivalent to $E_{B-V}=0.54$
\citep{2005AJ....129.1396H}.  However, we can estimate that the
average extinction experienced by an object at this position in M31
would be $A(r')=0.7$ magnitudes, equivalent to $E_{B-V}=0.25$
\citep[][see above]{2005PhDT.........2D}. 

We also calculate the position of a quiescent RS Oph system on this
diagram.  We use the LT $V$ and $i'$ luminosities of RS Oph in the
time range of 400-1300 days following the 2006 outburst
\citep[see][for days 400-600]{2009ASPC..401..203D} to estimate the mean quiescent magnitudes,
$<V>=11.03\pm0.03$, $<i'>=9.34\pm0.02$.  These magnitudes were then corrected for
the extinction towards RS Oph 
\citep[$E_{B-V}=0.7\pm0.1$,][]{1987rorn.conf...51S} and the distance to RS Oph \citep[$d=1.6\pm0.3$\ kpc,][]{1987rorn.conf..241B}.  The Sloan-$i'$ flux was transformed to the Johnson-Cousins system, the system
was placed at the distance of M31 and reddened by an amount equal to
the extinction towards that galaxy.  We find that the expected mean 
quiescent magnitude of an RS Oph-like system in M31 (without any
internal extinction) is $<I>=21.0\pm0.5$ with a color of
$<V-I>=1.3\pm0.4$.  Further correcting for the expected average internal
M31 extinction yields, $<I>=21.5\pm0.5$, $<V-I>=1.8\pm0.4$. We note that these values of quiescent magnitudes and colors include contributions from other sources than the secondary (e.g. any accretion disk).








\section{Conclusions}



M31N 2007-12b shows several characteristics consistent with it being a recurrent nova. These include the rapidity of its optical decline, extremely high ejection velocities and early emergence of its SSS phase. The early post-outburst optical spectrum also shows some similarities to that of RS Oph, but most closely resembles that of the proposed RN V2491 Cyg. Furthermore, we have found a coincident pre-outburst stellar source from archival {\it HST} observations that resides in the same region of the color-magnitude diagram as RS Oph. If this is indeed the quiescent nova system, this is the first time that this has been identified in a nova in M31. This finding also implies an outburst amplitude of $\Delta V \simeq 8.5$ mag, very similar to that given by \cite{jur08} for nova V2491 Cyg, although around 1 mag greater than that for RS Oph. The observed flux from the SSS detected in M31N 2007-12b is also more consistent with that of the short-lived peak at around 40 days in V2491 Cyg than that of the SSS in RS Oph. On the other hand, the secondary maximum reported in the light curve of V2491 Cyg at around $t = 15$ days is not apparent in our LT data for M31N 2007-12b.


Among the Galactic RNe, both U Sco and
RS Oph sub-types have been proposed as progenitors of Type Ia SNe as $M_{\rm WD}
\sim ~$M$_{\rm Ch}$ and it has been concluded that there is a net
accumulation of mass on the WD over time \citep{kah99,hac07}.
Determination of the true nature of Type Ia progenitors is of course a
very important quest for contemporary astrophysics, but still remains
a controversial area. recurrent novae have been one of the favored
systems, but  there are likely to be more problems in explaining the lack of H in SNIa spectra for RS Oph-like than for U Sco-like RNe. Certainly, the paucity of Galactic examples remains a hindrance to
further progress. We have shown that it is now possible to
identify RNe in M31, and even to determine their
sub-type, via a suitable set of complementary
observations.   However, from our experience we caution that identifying RNe from positional (near) coincidence of two or more outbursts can be precarious \citep[e.g., see][]{sha09b}.  All RN candidates should be thoroughly explored through precise astrometry of the original images, where available.  Furthermore, ambiguities of distance and host stellar
population are negated for novae in M31, and the soft X-ray absorbing column is low, compared to their Galactic
counterparts. Thus the prospects are good for extending our studies of
RNe, and in particular exploring any relationship to supernovae, from
the Milky Way to potentially a much larger and better-defined sample
of objects in the Andromeda galaxy.

\acknowledgments

We are grateful to M. Shetrone and A. Westfall for help with
  obtaining the HET spectroscopy and to S. Starrfield for suggesting use of the HST observations. The authors also wish to thank two anonymous referees for very helpful comments on the initial versions of the manuscript. The Liverpool Telescope is operated on the island of La
Palma by Liverpool John Moores University in the Spanish Observatorio
del Roque de los Muchachos of the Instituto de Astrofisica de Canarias
with financial support from the UK Science and Technology Facilities
Council.  AWS acknowledges support through NSF grant AST-0607682.

{\it Facilities:} \facility{Hubble Space Telescope},  \facility{Hobby Eberley Telescope}, \facility{Liverpool Telescope}, \facility{Swift}.

\clearpage

\clearpage

\clearpage

\clearpage

\end{document}